\documentclass[english]{IEEEtran}
\usepackage{color}
\usepackage{mathtools}
\usepackage{amsmath}
\usepackage{amssymb}
\usepackage{tabularx}
\usepackage{epsfig}
\usepackage{graphicx}
\usepackage[T1]{fontenc}
\usepackage{epstopdf}
\usepackage{cite}
\usepackage{float}
\usepackage{amsmath}
\usepackage{url}

 \usepackage{pdfpages}

\usepackage{amsmath}

\usepackage{algorithm}
\usepackage{algpseudocode} 
\begin{document}

\title{Decentralized Identifier-based
Privacy-preserving Authenticated Key  Exchange  Protocol for Electric Vehicle Charging in Smart Grid}

\author{
        Rohini Poolat Parameswarath,  Prosanta Gope, \textit{Senior Member, IEEE} and Biplab Sikdar, \textit{Senior Member, IEEE}
        
      \thanks{R.P Parameswarath and B. Sikdar are with the Department of Electrical and Computer Engineering, College of Design and Engineering, National University of Singapore, Singapore. (Email: rohini.p@nus.edu.sg, bsikdar@nus.edu.sg)\\
        Prosanta Gope is with the Department of Computer Science, University of
        Sheffield, United Kingdom. (E-mail: p.gope@sheffield.ac.uk)\\
    
        }

}

\maketitle

\begin{abstract}

The popularity of Electric Vehicles (EVs) has been rising across the globe in recent years. Smart grids will be the backbone for EV charging and enable efficient consumption of electricity by the EVs. With the demand for EVs, associated cyber threats are also increasing. Users expose their personal information while charging their EVs, leading to privacy threats. This paper proposes a user-empowered, privacy-aware authenticated key exchange protocol for EV charging in smart grid. The proposed protocol is based on the concept of Decentralized Identifier (DID) and Verifiable Credentials (VCs). The use of DIDs empowers users by helping them to have complete control over their identities. The charging station and the user verify that the other party is legitimate before proceeding with the charging services using VC. Key recovery is another issue we address in this paper. A method to recover lost keys is incorporated into the proposed protocol. We present formal security proof and informal analysis to show that protocol's robustness against several attacks. We also provide a detailed performance analysis to show that the proposed protocol is efficient.

\end{abstract}


\section{Introduction}

There has been an increased interest in the use of Electric Vehicles (EVs) in recent years. While the registrations of the conventional cars fell in 2020, electric car registrations increased in 2020 despite the Covid pandemic \cite{IEA}. There are various reasons for this trend. Traditional vehicles run on internal combustion engines that burn hydrocarbons, resulting in air pollution and greenhouse effect. Since EVs use electricity, they result in almost no air pollution \cite{li2020privacy} compared to the traditional gasoline-powered vehicles and are more environmentally friendly. Hence, authorities across the globe encourage the use of EVs \cite{antoun2020detailed} and provide incentives to car users to make the switch to EVs. 

A smart grid is an electrical grid that allows the monitoring of power flow from generation to consumption, and load regulation to match power generation. Advanced technologies are employed in smart grids to improve power generation and transmission. Coupling EV charging with smart grid infrastructure has several advantages. 
As EVs need a significant amount of electricity for charging, they add load to the power distribution network. 
Smart grid infrastructure enables the efficient consumption of electricity by EVs. EVs support the concept of vehicle-to-grid (V2G) technology, where they take electricity from the grid while charging and give it back to the power grid from the battery when demand is high \cite{roman2019pairing}. As a result, EVs can also help lessen the smart grid's burden during periods of high power demand. 
Smart grid technologies also allow EVs to be identified by charging stations, and the electricity consumed can be automatically billed to the owner's account. The participants in the EV charging ecosystem are Utility Service Providers (USPs), Charging Stations (CSs), and vehicles. The USP generates and distributes power to CSs \cite{gope2019efficient}. 

Though EVs are better for the environment and paves the path to a sustainable transport system, security and privacy concerns have surfaced with the increased usage of EVs. An attacker may track where users charge their EVs, and gather personal details of the users. Tracking the activities of EV users allows the attacker to obtain their footprints. There have been studies on how an attacker can misuse such sensitive information for stalking or physical attacks \cite{Krumm2007127}, \cite{schilit2003wireless}. Marketing and advertising companies can leverage such information to send unsolicited advertisements. To address such privacy threats, this paper proposes a privacy-aware protocol based on Decentralized Identifier (DID) and Verifiable Credential (VC) for EV charging. To remain anonymous and to protect privacy, the use of pseudo-IDs instead of real identities of users has been proposed in literature \cite{gope2019efficient}. The pseudo-IDs have been assigned by a central authority to the users. The World Wide Web Consortium (W3C) DID Working Group has created a DID standard that allows the users to create and manage their identities. Thus, in the DID mechanism,  the users create and manage their IDs and they have complete control over their IDs.  They do not have to rely on a central authority to assign IDs. A trusted party can check the legitimacy of the user and sign credentials that other parties can verify digitally before providing services. VCs have been introduced for this purpose. When DID is combined with the concept of VC, any third party can verify the legitimacy of the user. As a result, by combining DID and VC, the proposed protocol ensures that users remain anonymous and their identity can be verified by others.

\subsection{Related Work}

To protect the communication between different entities in smart grids from attacks, many authentication schemes have been proposed. Researchers have been looking at the security and privacy issues in EV charging as well. The authors of \cite{8994200} presented a security evaluation of the EV charging framework. The cyber threats faced by different entities, together with the available defences available discussed in \cite{8994200}. Privacy threats that exist in EV charging infrastructure and recommendations to protect the privacy of users were presented in \cite{unterweger2022analysis}. A role-dependent privacy preservation scheme was proposed in \cite{role6684311} based on whether the EV is a customer or supplier of energy. In each role, the privacy concerns are different and addressed separately. An authentication scheme for V2G networks using elliptic curve cryptosystem and bilinear pairing was proposed in \cite{zhang2021efficient}. An identity-based sequential aggregate signed data based on homomorphic encryption was proposed in \cite{9273059}. In the scheme proposed in \cite{7997252} for EV charging, hashing and XOR operations are used. An authentication scheme when EVs 
roam from one network to another in V2G networks was proposed in \cite{Neetesh7414504}. An authentication scheme where the power grid ensures the confidentiality and integrity of the messages was proposed in \cite {Abdallah7485857}. However, their scheme does not provide mutual authentication. Without mutual authentication, one party is not sure whether the person on the other side is legitimate. A privacy-preserving V2G authentication scheme based on Public Unclonable Functions (PUFs) was presented in \cite{kaveh2020lightweight}. Another PUF-based authentication protocol for the V2G framework was proposed by \cite{Bansal9018072}. However, such protocols require the additional hardware, PUF. While charging EVs, another solution to protect users' privacy is to use pseudonyms rather than their real names. The authors of \cite{nicanfar2013robust} proposed an authentication scheme where  the pseudonyms of users are generated by the smart grid servers. CSs generate pseudonyms for EVs in the anonymity-based authentication proposed in \cite{kilari2016revocable}. A V2G authentication scheme with pseudo-identities of users was was proposed in \cite{gope2019efficient}. The pseudo-identities are generated by service providers.
The electricity suppliers assign pseudonyms in \cite {7007769}. In the authentication scheme in \cite{saxena2015lightweight},  certificate authorities issue pseudo-ids to the EVs. The pseudo-identities of vehicles are issued by third-party authorities in \cite{8809366}.

\subsection{Motivation}

We can see that several works have been proposed in the literature to achieve security and privacy in EV charging. However, none of them provides a user-empowered authentication for V2G. In the existing works, the pseudo-identities are managed by CSs or other servers where the IDs are created and stored on central servers. Because all sensitive information is stored on a central server in centralized systems, there is a risk of information leakage. There is also a risk of a single point of failure in such centralized systems.
To address the above issues, a user-empowered privacy-aware authenticated key exchange protocol for EV charging by combining DID and VC is proposed in this paper. The users do not need to rely on a central issuing authority for their IDs because DIDs are created and controlled by themselves. As a result, DIDs assist users in creating, managing, and controlling their identities without the assistance of a central issuing authority. The authors of \cite{UserEmpowered} presented an authentication scheme for EV charging that discussed the importance of user empowerment. However, they have not considered some key issues such as the usability problem and the unauthorized use of mobile devices. The usability problem refers to the scenario where an adversary has access to the mobile device of a user for a short duration and tries to delete the private key before being noticed by the user. This will result in the loss of the private key and identity loss of the user. The proposed protocol addresses this problem and incorporates a solution for key recovery. To prevent the unauthorized use of the mobile device of a legitimate user for EV authentication, biometric verification to access the mobile device is also included in the proposed protocol. Instead of the Rivest–Shamir–Adleman (RSA) algorithm used in \cite{UserEmpowered}, the proposed protocol is based on Elliptic Curve Digital Signature Algorithm (ECDSA) that uses small key size.

\subsection{Our Contributions}

This paper makes the following key contributions:

\textbf{1. A \emph{new} user-empowerment based privacy-aware authenticated key exchange protocol based on DID and VC:} We propose a new protocol for EV charging in smart grids by combining DID and VC. The proposed protocol enables users to create, manage, and control their IDs, which not only empowers them but also helps to preserve privacy. The decentralized nature of DID eliminates the dependency on a centralized system. The USP issues VCs to legitimate users. The CS allows users who show the VCs issued by the USP. The proposed protocol provides several important security properties and establishment of a secure session key after completing successful authentication.

\textbf{2. Key recovery employing secret sharing technique:} The proposed protocol includes a key recovery mechanism so that the user's private key can be stored securely and recovered if it is lost. We employ Shamir's $(k,n)$ threshold secret sharing scheme \cite{shamir1979share} for private key recovery.  

\textbf{3. Two-factor authentication:} The user inputs his/her biometrics (e.g., fingerprint) to access the mobile device before initiating the authentication process. This biometric verification on the mobile device ensures the user's legitimacy. Even if an adversary gets the mobile device, he/she cannot use it for EV charging authentication as the biometric verification will fail.

\textbf{4. Security analysis:} We provide formal security proof based on \cite{abdalla2005password} and informal security analysis to demonstrate the proposed scheme's robustness against common attacks. 

\textbf{5. Performance analysis:} We provide  detailed performance analysis and comparison of the performance of the proposed method with other existing methods. 

The rest of the paper is organized as follows. In Section II, we introduce the preliminaries. In Section III, the system and the adversary models are presented. We present the proposed protocol in Section IV. The formal security proof is presented in Section V and the informal security analysis is presented in Section VI. We provide performance evaluation and comparison with other schemes in Section VII. The conclusions are drawn in Section VIII.

\section{Preliminaries} \label{Prelims}

This section begins with a discussion of the concepts of Decentralized Identifier and Verifiable Credentials. After that, we discuss the cryptographic and mathematical concepts behind the proposed protocol.

\begin{figure}[t!]
    \centering
    \includegraphics[width=0.23 \textwidth]{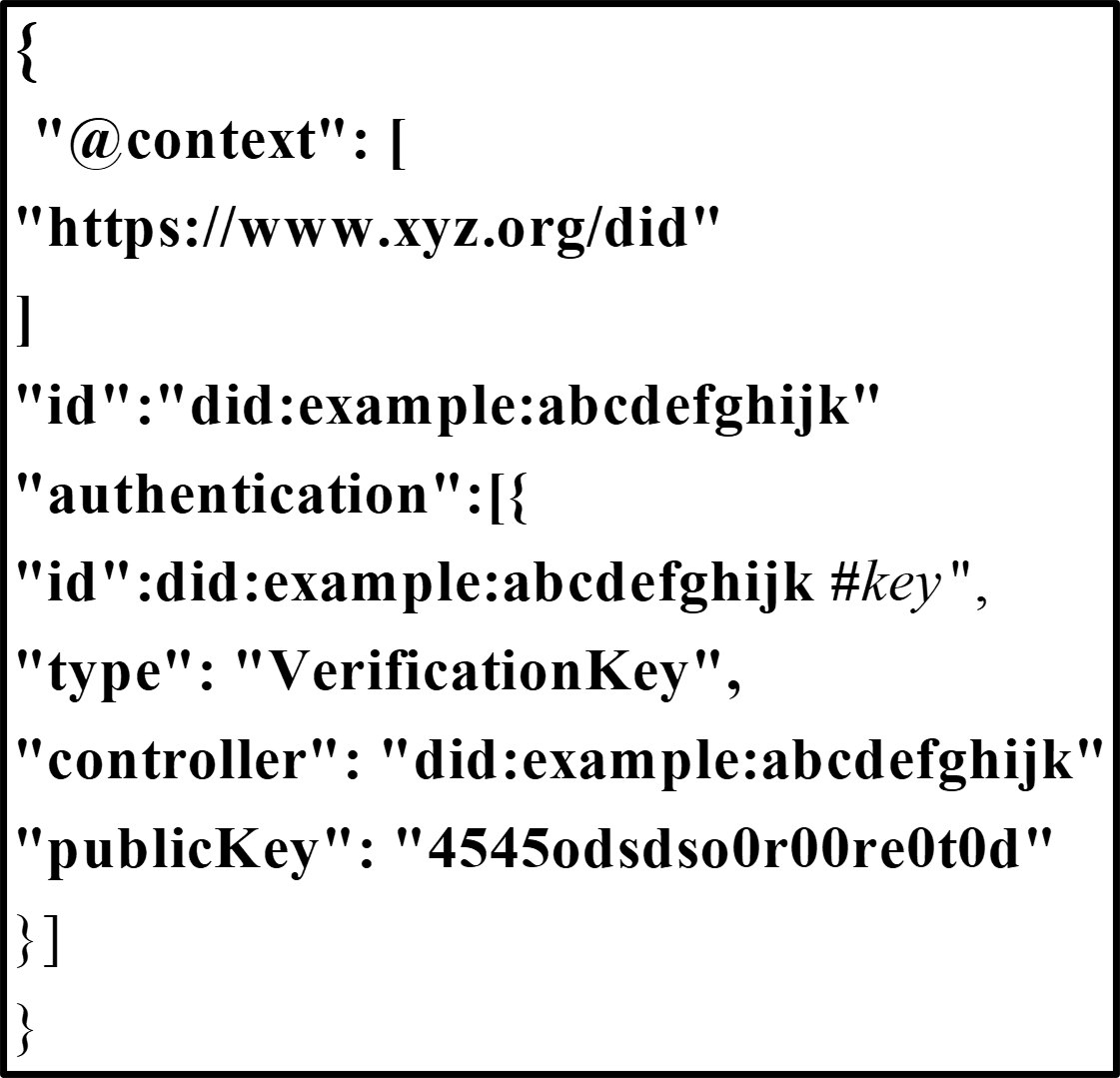}
    \caption{An example of a DID document in JSON representation.}
    \label{fig:DID}
\end{figure}

\textbf{Decentralized Identifier:} DID is a type of identifier which is created and controlled by the user. As no centralized authority is required to issue a DID, DID empowers users to have complete control and ownership over their IDs. W3C has developed the standard for DID \cite{W3cwebsite}. A person can have different DIDs to ensure that they are not tracked by correlating their activities. A DID resolves to a DID document. The DID document contains information about the DID owner, such as the owner's public key. The DID document can be stored on a public ledger such as a blockchain. A DID is of the form $\textbf{did:<DID method>:<method-specific identifier>}$. The DID method is a reference to the underlying distributed ledger. The method-specific identifier resolves the DID to the DID document on the ledger. An example of a DID document for the DID $\textbf{did:<example>:<abcdefghijk>}$ is shown in in Figure \ref{fig:DID}.

\begin{figure}[h!]
    \centering
    \includegraphics[width=0.24 \textwidth]{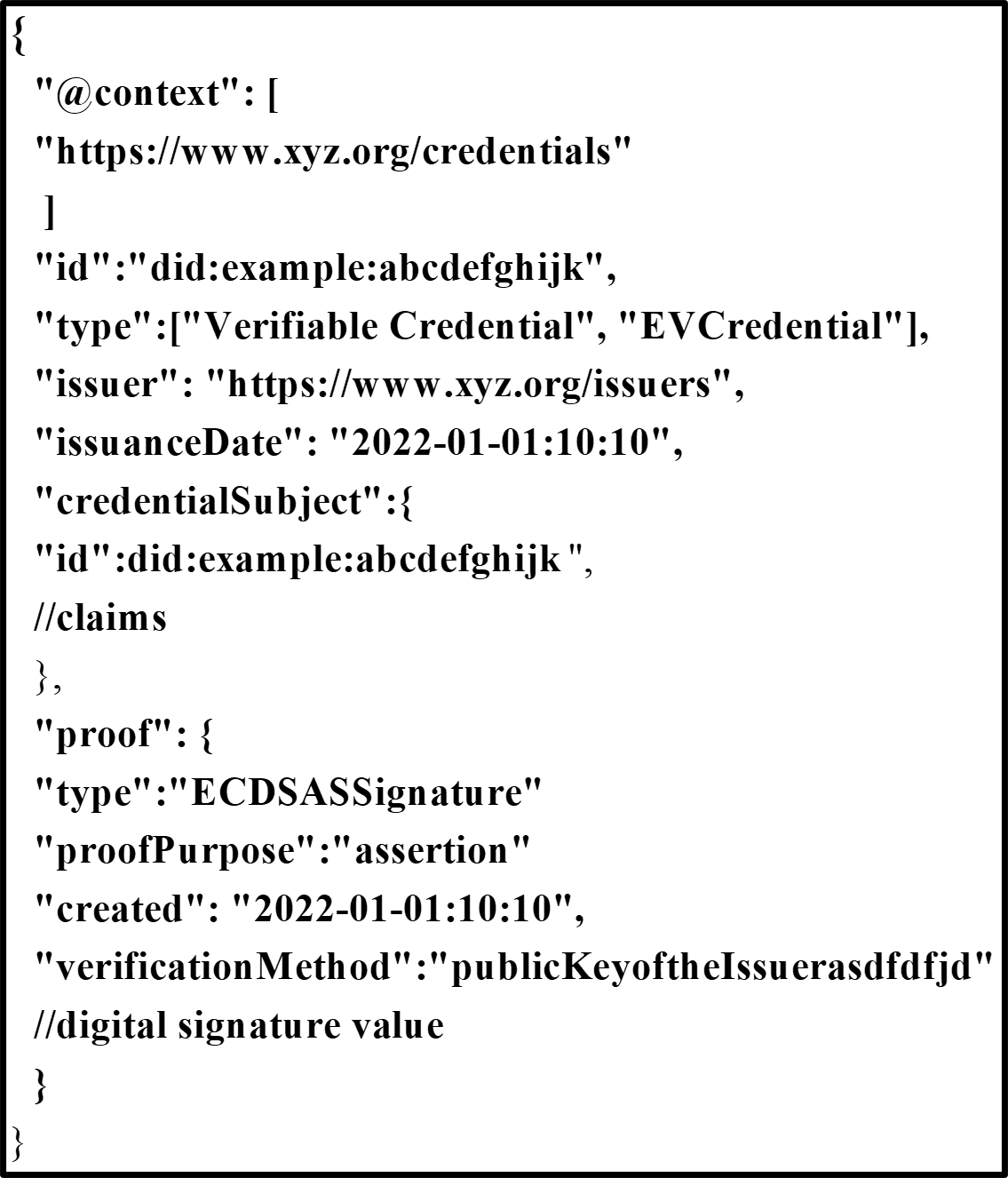}
    \caption{An example of a Verifiable Credential.}
    \label{fig:VC}
\end{figure}

 \textbf{Verifiable Credential:} A VC is a set of claims that can be verified using cryptographic techniques such as digital signatures \cite{W3cwebsite}. The users create their DIDs themselves. The concept of VC can be used to ensure the claims of the user by another party. The VC ecosystem consists of an \textit{Issuer, a Holder, and a Verifier}. A trusted \textbf{Issuer} issues credentials about the VC \textbf{Holder} and signs it digitally. Another party, \textbf{Verifier}, can verify the statements about the \textbf{Holder} \cite{W3cwebsite}, \cite{Microsoft}. Since cryptographic techniques are used, VCs are tamper-resistant and can be verified digitally by others. An example of a VC is shown in Figure \ref{fig:VC}. Verifiable Presentation (VP), defined by W3C, enables a user to present the VC to a verifier. VP can be presented using Zero-Knowledge Proof. 

\begin{figure} [t]
    \centering
 \includegraphics[width=.48 \textwidth]{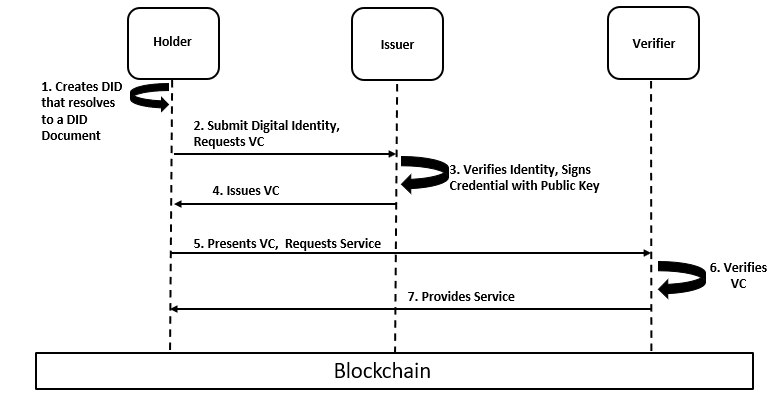}
 
    \caption{DID and Verifiable Credential.}
    \label{fig:DIDandVC}
\end{figure}

 Figure \ref{fig:DIDandVC} illustrates the workflow of the interaction between DID and VC \cite{W3cwebsite}. A DID subject creates his/her own DID. The DID resolves to a DID document as shown in Figure \ref{fig:DID} that resides on decentralized systems such as blockchain. A trusted \textit{Issuer} verifies the user's identity, then signs and issues a VC to the user. To get services, users disclose their DIDs to the \textit{Verifier}. Then, the user presents the signed VC. After that, the \textit{Verifier} verifies the signed VC by using the public key of the Issuer.
 
 \textbf{Zero-Knowledge Proof (ZKP):} The ZKP enables a person (Prover) to present the knowledge of a value to someone else (Verifier) but without revealing any other information. Non-interactive zero-knowledge proof (NIZKP) is a type of ZKP that requires reduced interaction between the two parties. A common reference string is shared between the Prover and the Verifier to achieve ZKP \cite{groth2016size}, \cite{de2022leveraging}, \cite{salleras2021zpie}. A ZKP consists of the algorithms presented in Table \ref{table:zkp} \cite{groth2016size}. In the algorithms mentioned in Table \ref{table:zkp}, a relation generator returns a polynomial-time decidable binary relation $R$ for a security parameter $\lambda$. For pairs $(st,w) \in R$, $st$ is the `statement' and $w$ is the `witness'. $crs$ is a common reference string and $td$ indicates the simulation trapdoor.
There are three properties for ZKPs \cite{groth2016size}. They are completeness, Zero-Knowledge, and soundness.

\textbf{Completeness:} An honest prover can convince a true statement to an honest verifierL
\begin{multline*} 
Pr [(crs, td)] \leftarrow Setup(R); \pi \leftarrow Prove(R, crs, st, w):\\ \mathit{Verify}(R, crs, st, \pi) = 1] =1.
\end{multline*}

\begin{table*}[t!]
		\begin{center}
		\caption{ZKP Algorithms}
			\begin{tabular}{|c|c|}
			\hline
			\textbf{Algorithm} & \textbf{Description} \\
			\hline
				$(crs, td )$ ← $Setup(R)$  & For the relation $R$,   $crs$ and $td$ are generated. \\
						\hline
		  	
		  		$\pi$← $Prove(R, crs, st, w)$
				     	
	    	 & This algorithm takes  $crs$ and $($st$, w)\in R$ as inputs and returns an argument $\pi$. \\
				
				\hline
				$0/1$ ← $Verify(R, crs, st, \pi)$ & This algorithm takes $crs$, $st$ and $\pi$ as inputs and returns 0 or 1 for reject or accept, respectively. \\
				
				\hline
				$\pi$ ← $Sim(R, td, st)$ & The simulator takes $td$ and $st$ as inputs 
               and returns $\pi$. \\
	
          	\hline
	   	\end{tabular}
			
			\label{table:zkp}
		\end{center}
	
	\end{table*}	

\textbf{Zero-Knowledge:} The proof does not reveal anything other than the truthfulness of the statement.
 For all $\lambda \in N$,
 $(R, z) \leftarrow R(1^\lambda )$, 
 $(st, w) \in R$ and adversary $A$, we can write:
\begin{multline*} 
Pr [(crs, td)] \leftarrow Setup(R); \pi \leftarrow Prove(R, crs, st, w):\\ A(R, z, crs, td, \pi) = 1]\\
=Pr [(crs, td)]\leftarrow Setup(R); \pi \leftarrow Sim(R, td, st)\\: A(R, z, crs, td, \pi) = 1].
\end{multline*}

\textbf{Soundness:} A prover cannot prove a false statement to the verifier:
\begin{multline*} 
Pr [(R,z) \leftarrow R(1^ \lambda); (crs, td) \leftarrow Setup(R);\\ 
 (st, \pi) \leftarrow
A(R, z, crs):  st \notin L_R \; \land \mathit{Verify}(R, crs, st, \pi)] =1 .
\end{multline*}

  \textbf{Asymmetric Cryptography and ECDSA:}
  Asymmetric cryptographic technique can be used to encrypt or sign data using a pair of keys (public and private). The private key is solely known by the owner and is not shared with anyone else. Others have access to the public key.  
   Elliptic-Curve Cryptography (ECC) \cite{koblitz1987elliptic} is a public key cryptography technique. ECC is considered to be secure due to the difficulty of the Elliptic Curve Discrete Logarithm Problem (ECDLP) and Elliptic Curve Decisional Diffie–Hellman problem (ECDDHP) \cite{20142517841778}, \cite{ECDLP} that are defined below.

\textbf{Elliptic-Curve Discrete Logarithm Problem (ECDLP):} Let $p$ be a prime number and $q = p^n$. Let $E$ be an elliptic curve over a finite field $F_q$. If points $P, Q \in E(F_q)$ are given, ECDLP is the computational problem to find the integer $a$, if it exists, such that $Q = aP$.

\textbf{Elliptic-Curve Decisional Diffie-Hellman Problem (ECDDHP):} Given $P, aP, bP, Q \in E(F_q)$, ECDDHP is to determine if $Q = abP$.

 ECDSA is a popular algorithm used in asymmetric cryptography based on ECC. In the proposed protocol, we use ECDSA. Cryptosystems based on elliptic curves use small key size. They have low memory usage and need less processor resources. Due to these properties, they are ideal even for resource-constrained devices. ECDSA signature generation and verification algorithms are shown in Algorithms \ref{alg:ECDSASign} and \ref{alg:ECDSAVerify}, respectively \cite{ECDSACryptobook}. The public key is calculated as $pubKey = priKey \times G$. In Algorithm \ref{alg:ECDSASign}, a message $m$ is signed with $priKey$ to get the signature $( r, s)$. In Algorithm \ref{alg:ECDSAVerify}, the message $m$, the signature, and the public key are the inputs and based on whether the signature is valid or not, it is accepted or rejected.

\begin{algorithm}
\caption{ECDSA Signature Generation, $Sign_{ECDSA}(m, priKey)$ -> $( r, s )$}\label{alg:ECDSASign}
\begin{algorithmic}

\State Calculate $h= hash (m)$
\State Generate a random number $k$ in the range $[1..n-1]$
\State Calculate $R= r \times G$ and its x-coordinate $r=R.x$

\State The signature $s = k^{-1} \times (h + r\times priKey) (\mod n)$

\State Signature is $(R, s)$

\end{algorithmic}
\end{algorithm}

\begin{algorithm}
\caption{ECDSA Signature Verification,\\ 
$Verify_{ECDSA}(m, pubKey, Signature)$-> $(Accept, Reject)$} \label{alg:ECDSAVerify}
\begin{algorithmic}

\State Calculate $h= hash (m)$
\State Calculate  $s^{'}= s^{-1} (\mod n)$
\State Calculate $R^{'} = (h \times s^{'}) \times G + (r \times s^{'}) \times pubKey$ and its x-coordinate $r^{'}=R^{'} \times x$

\If{$r^{'}	= r$} 
  \State $Accept$
\Else
  \State $Reject$
\EndIf
\end{algorithmic}
\end{algorithm}

 \textbf{Key Recovery and Shamir's Secret Sharing:} 
 The DID ecosystem is built on a public key infrastructure where users need to manage their private keys. Users should have a secure backup of their private keys. Hence, key possession and recovery of the lost key are crucial in maintaining trust \cite{paper2}. Key recovery refers to methods for securely backing up private keys so that they can be recovered if the private key is lost in events such as the device containing the private key is damaged. We use the famous Shamir's threshold scheme \cite{shamir1979share} for private key recovery. The key is encoded into a polynomial. After that, it is divided into pieces. Using polynomial interpolation, the key can be computed with a threshold value of the pieces.

In this method, a key $D$ is split into $n$ pieces $D_1, D_2, \cdots, D_n$ such that:

(1) $D$ can be computed from any $k$ (threshold value) or more pieces and 

(2) $D$ cannot be calculated when there are only $k-1$ or fewer pieces.

To divide $D$ into $k$ shares, the holder of the key selects a a random $k-1$ degree polynomial as:

\begin{equation}
f(x)= D+a_1(x)+ \cdots +a_{k-1}(x)^{k-1}\label{pre}.
\end{equation}

In (\ref{pre}), $a_1, a_2, \cdots,  a_{k-1}$ are random polynomial coefficients. After that, the $n$ values are evaluated as $D_1=f(1)$, $D_2=f(2)$, $\cdots$, $D_n=f(n)$. With any subset of $k$ of these $n$ values, the key can be calculated as
 \begin{equation} \label{keyrecovery}
D = \Sigma_{j=1}^k f(i_j)  \Pi_{j\neq m} \frac{i_j}{i_j - i_m}. 
 \end{equation}
 
By knowing $k-1$ of these values, $D$ cannot be calculated. This method of secret sharing can be employed for key recovery.

\section {System and Adversary Model}

\subsection{System Model}
\begin{figure} [h!]
    \centering
    \includegraphics[width=0.42\textwidth]{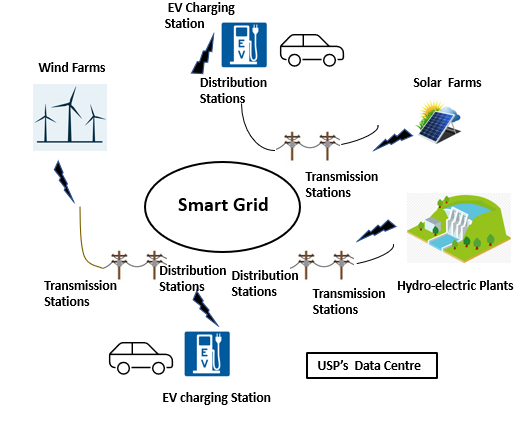}
    \caption{System model.}
    \label{fig:framework}
\end{figure}

The EV charging system model considered in this paper is illustrated in Figure \ref{fig:framework}. The USP, the CS and the users are the participants in this model. The USP is in charge of power generation, distribution to different CSs, and also data management. Electricity is generated from different sources such as wind farms, solar farms, and hydroelectric plants. Then, it is transmitted and distributed to the CSs. The charging rate at a CS depends on its location. The users use their mobile devices (MDs) to connect to the Internet. All the participants in the model communicate with each other through the Internet. 

\subsection{Adversary Model}

During authentication, the participants exchange messages over an insecure channel, the Internet over which attackers may launch multiple attacks. According to Dolev-Yao threat model, an adversary can listen, modify or delete the messages sent between different parties. Hence, the following threats exist against the system model:

\textbf{Data Modification Threat:} Since the attacker is able to eavesdrop, edit or delete the messages exchanged, there is a threat of data modification. A legitimate user may be denied access due to such data modification. 

\textbf{Unauthorized Access Threat:} The attacker can capture legitimate messages and replay it later to get authenticated as a legitimate user. An attacker may impersonate a registered user to charge the vehicle.
 \begin{figure*}[t]
    \centering
   \includegraphics[width=0.65 \textwidth]{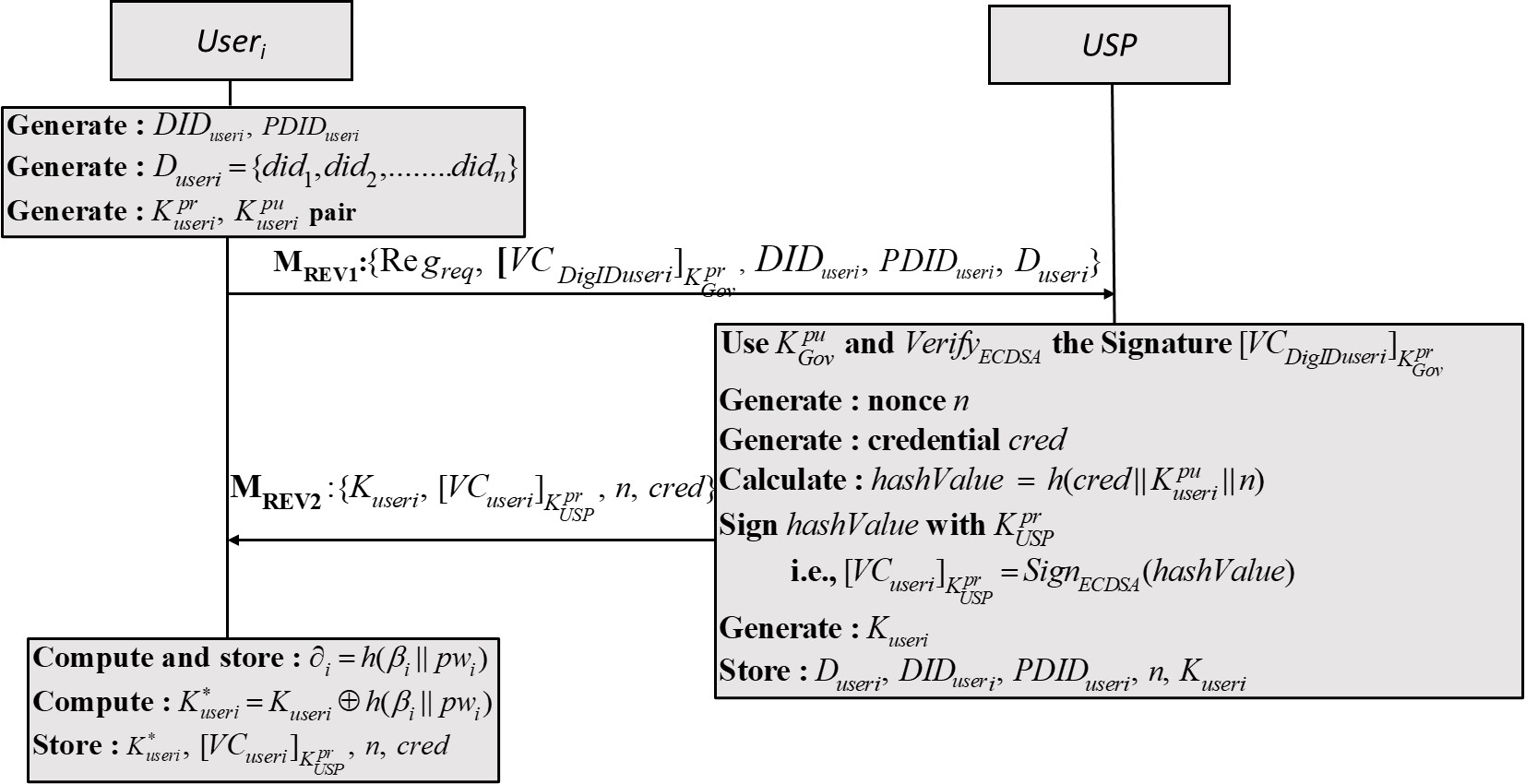}
 
    \caption{EV registration phase.}
    \label{fig:EVregistration}
\end{figure*}

\textbf{Privacy Threat:}An adversary may gether information about the user's presence in a location at a particular time, his/her daily routines and trajectory, and charging pattern, from the data related to the user's EV charging. 

\textbf{Stealing of Mobile Device Threat:} The adversary has the capability to steal the mobile device of a legitimate user. Then, the adversary may use it to initiate EV authentication while impersonating the user. 

\textbf{Loss of Private Key Threat:} We also consider the scenario of `Usability Problem' where the adversary has access to the mobile device of a legitimate user for a couple of minutes, and tries to delete the private key before being noticed by the owner of the device. This will result in loss of private key and identity loss of the user. 

\textbf{Location Forgery Threat:} The USP can be trusted in the proposed model, but the the other two parties may not be trustworthy. A dishonest user may provide the incorrect location area identifier (LAI) of the CS where he/she wishes to charge EV in order to pay a lower price than the actual rates. A dishonest CS may also provide an incorrect LAI in order to obtain higher EV charging rates from users. As a result, location forgery threat also exists.

\section {Proposed Authentication Protocol}

The privacy-aware authenticated key exchange protocol for EV charging is presented in this section. The participants in this protocol are the USP, the user, and the CS. The USP issues a VC to the user, and the CS verifies the VC of the user before providing charging services. The three phases in the proposed protocol are: \emph{system setup and registration}, \emph{authentication}, and \emph{key recovery}. The system setup and registration phase is  only performed once. Before charging their vehicles, registered users are required to go through the authentication process. The key recovery phase is executed if the user wants to retrieve a lost private key. The proposed protocol is built on the DID and VC standard given in \cite{W3cwebsite}. The high-level workflow of the EV charging protocol can be explained based on Figure \ref{fig:DIDandVC}. The user creates his/her DID that resolves to a DID document. Then, the user submits a signed digital identity to the USP for verification. The USP verifies the user’s digital identity. If the digital identity verification is successful, the USP generates a credential. Then, the USP signs it with its public key and issues the signed VC to the user. The user generates the ZKP of the signed VC and presents it to the CS together with the charging request. The CS verifies the signature using the USP’s public key. After successful verification, the CS provides service to the user. The notations used in the proposed authentication protocol are presented in Table \ref{table:notations}.

\begin{table}[t!]
		\begin{center}
		\caption{Notations}
			\begin{tabular}{|l|l|}
				\hline

			Symbol & Description \\
			
							\hline
				$G $ & Generator Point \\
			
				\hline
				$a_1,a_2,..,a_{k-1} $ & Polynomial Coefficients for key recovery \\
			   	\hline

			$DID_{x}$ & DID of the entity x \\

				\hline
				$PDID_{useri}$ & Pseudo-identity of $User_i$ \\
				
				\hline
				$D_{useri}$ & Set of unlinkable shadow identities of $User_i$ \\

				\hline
				$K^{pr}_{x} $ & Private key of the entity $x$ \\
				\hline
					$K^{pu}_{x} $ & Public key of the entity $x$ \\		
	
				\hline
				
					$MD_{useri}$ & Mobile device of $User_i$ \\
				\hline

					$r$ & Nonce generated by the USP \\
				
					\hline

				$K_{useri}$ & Secret key between USP and $User_i$ \\
				
					\hline
				
				$K_{CSj}$ & Secret between USP and  $CS_j$ \\
				
					\hline

				$[VC_{useri}]_{K^{pr}_{USP}}$  & VC of $User_i$ signed by the USP  \\
				
				\hline
	  
		    	$SK$ & Session key  \\
				
				\hline

				$\parallel $ & Concatenation operation  \\
				\hline
					$\oplus $ & XOR operation  \\
				\hline
				
				$h(X)$ & Hash of X \\
				\hline

			\end{tabular}
			
			\label{table:notations}
		\end{center}
	
	\end{table}

\subsection{Assumptions}

We assume that a secure channel is used for communication between the parties in the registration phase. We assume that the USP can be trusted. The CS and the USP do not collude to learn additional information about the users. The users have to be authenticated  before charging their EVs each time, even if they have charged from the same charging station previously. The proposed protocol is also applicable in shared vehicle environments. In this scenario, each user must be a valid registered user. Following that, users must use their own VCs to get authenticated.

\subsection{System Setup and Registration Phase}

\subsubsection{System Setup} The system setup consists of the following steps:

\textbf{Step 1:} The USP's DID $DID_{USP}$ is stored on the blockchain. The USP generates its private key $K^{pr}_{USP}$ as a random integer. The public key is generated through ECDSA key generation function. Let $G$ be the generator point. The public key $K^{pu}_{USP}$ is a point on the elliptic curve, calculated by the Elliptic Curve (EC) point multiplication as $K^{pu}_{USP} =K^{pr}_{USP} \times G$. The USP's public key is stored in its DID on the blockchain. 

\textbf{Step 2:} $User_i$ generates a DID ($DID_{useri}$) and stores it on the blockchain, following which $User_i$ generates a pair of private ($K^{pr}_{useri}$) and public ($K^{pu}_{useri}$) keys using the ECDSA key generation function similar to the key generation mentioned in Step 1. The user stores $K^{pr}_{useri}$ in the digital wallet on the mobile device $MD_{useri}$. The public key $K^{pu}_{useri}$ is stored in the user's DID document on the blockchain.

\textbf{Step 3:} Similarly, the charging station, $CS_j$ (the Verifier), generates its DID ($DID_{CSj}$) and stores its public key immutably in its DID document on the blockchain.

\subsubsection{Registration} In this phase, the registration of the electric vehicles and the charging stations with the USP takes place.

\begin{figure*}[t]
    \centering
    \includegraphics[width=0.9\textwidth]{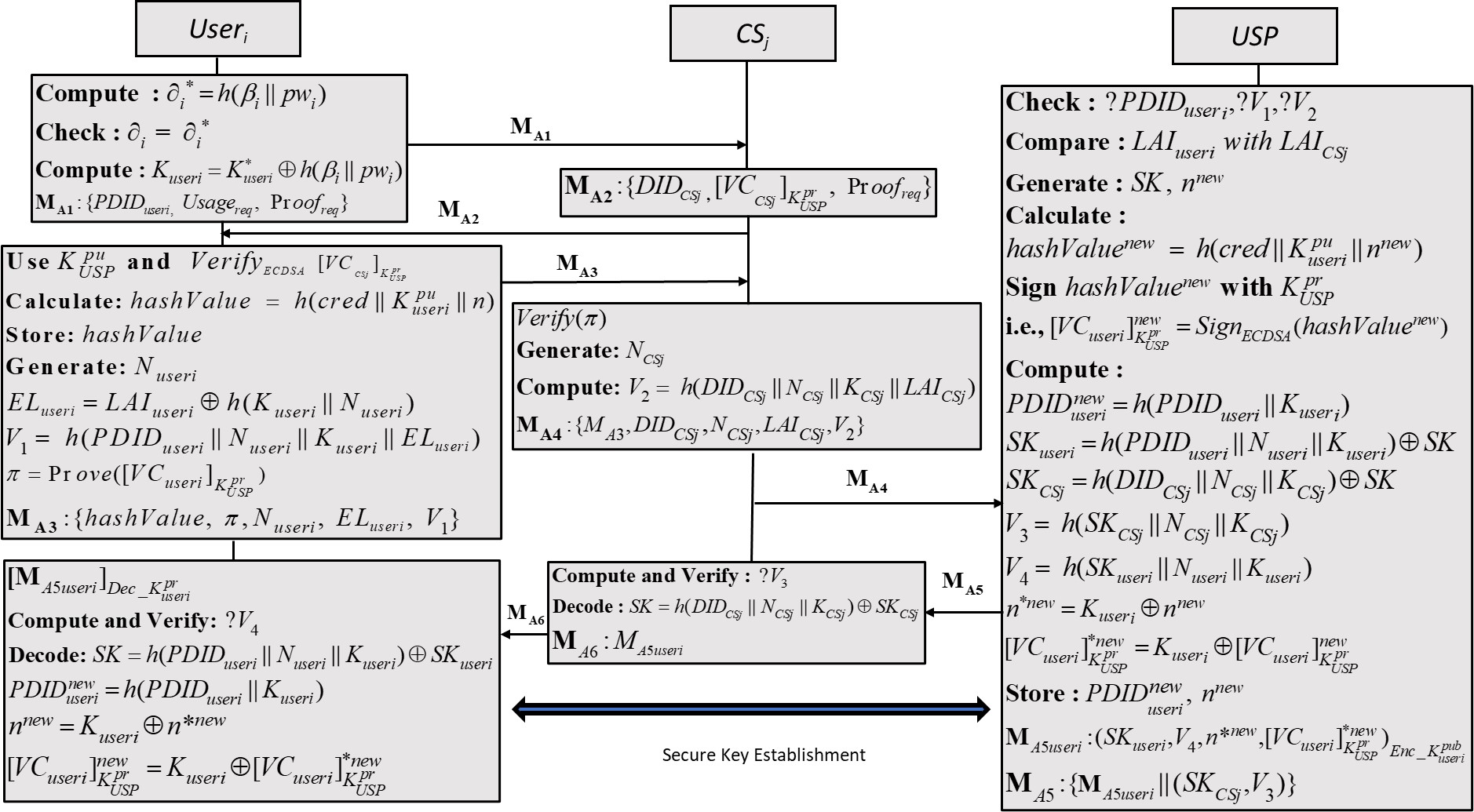}
    \caption{Authentication phase.}
    \label{fig:Auth}
\end{figure*}

\textbf{EV Registration:} The registration of the EV consists of the following steps:

\textbf{Step 1:} $User_i$ generates a pseudo-identity $PDID_{useri}$ and a set of unlinkable shadow identities $D_{useri}$ which will be used later to restore synchronization with the USP if required. The user holds ${[VC_{DigIDuseri}]_{K^{pr}_{Gov}}}$, the VC of his/her digital identity issued by a trusted party (e.g., a government agency). Then, the user generates a message $M_{REV1}$ with a registration request, $DID_{useri}$, the pseudo-identity $PDID_{useri}$, $D_{useri}$, and ${[VC_{DigIDuseri}]_{K^{pr}_{Gov}}}$. After that, $User_i$ sends $M_{REV1}$ to the USP.

\textbf{Step 2:} The USP needs to authenticate the user. The USP verifies the user's digital identity using ECDSA's signature verification algorithm as mentioned in Algorithm \ref{alg:ECDSAVerify}. After that, the USP generates a nonce $n$ and a credential $cred$. After that, $cred$, the user's public key, and $n$ are combined and its hash value $hashValue$ is calculated. Then, the USP signs $cred$ with its private key using ECDSA's signature generation algorithm as mentioned in Algorithm \ref{alg:ECDSASign}. For that, the USP generates a random number $k$ in the range $[1,m-1]$. The random point $R$ and its x-coordinate are calculated as: 

 \begin{equation}
 R = k \times G; 
 r = R.x.
 \end{equation}

 Then, the USP calculates the signature as:
 
  \begin{equation}
   k^{-1} \times (hashValue + r\times K^{pr}_{USP})(\mbox{mod } m). \label{step3_signature}
  \end{equation}

The verifiable credential $[VC_{useri}]_{K^{pr}_{USP}}$ of $User_i$ given in (\ref{step3_signature}) indicates that $User_i$ is a valid registered user. It can be verified using the corresponding public key $K^{pu}_{USP}$ of the USP. Thus, the USP certifies the user by signing the hash value with $K^{pr}_{USP}$ using ECDSA's signature generation algorithm as mentioned in Algorithm \ref{alg:ECDSASign}.
 
The USP generates a key $K_{useri}$ and then stores $DID_{useri}$, $PDID_{useri}$, set of pseudo-IDs, and $K_{useri}$ in its database for future communication with the user. After that, the USP generates a message $M_{REV2}$ with $K_{useri}$, $cred$, $n$, and the signed VC and sends it to the user.

\textbf{Step 3:} The user, $User_i$, receives $M_{REV2}$ from the USP. Then, $User_i$ inputs his/her biometrics (e.g., fingerprint) $\beta_i$ and password $psw_i$. This stored input will be used during authentication phase to ensure the user's legitimacy. Then, $User_i$ computes $K_{useri}^* = K_{useri} \oplus h(\beta_i \parallel psw_i)$. After that, $User_i$ stores $K_{useri}^*$, $cred$, $n$, and $[VC_{useri}]_{K^{pr}_{USP}}$ in the digital wallet on $MD_{useri}$.

Figure \ref{fig:EVregistration} depicts the EV registration step in detail. .


\textbf{CS Registration:} The following are the steps involved in CS registration:

\textbf{Step 1:} Similar to EV registration, $CS_j$ registers with the USP and receives a verifiable credential $VC_{CSj}$ signed by the USP with its private key $K^{pr}_{USP}$. The USP produces a key ${K_{CSj}}$ and sends it to the CS. The CS stores ${K_{CSj}}$ to use during the the authentication phase.


\subsection{ Authentication Phase}

 The user $User_i$ must go through the authentication process in order to charge the vehicle at $CS_j$. $User_i$ also authenticates $CS_j$.

\textbf{ Step 1:} $User_i$ inputs his/her biometrics $\beta_i$ and password $psw_i$ into his/her mobile device, $MD_{useri}$. Then, $MD_{useri}$ computes $\delta'_{i} =h(\beta_i \parallel psw_i)$ and compares $\delta'_i$ with $\delta_i$ to verify $User_i$'s legitimacy. This biometric verification at the mobile device is to ensure the user's legitimacy. If the biometric verification is successful, $MD_{useri}$ calculates $K_{useri} = K_{useri}^* \oplus h(\beta_i \parallel psw_i)$. Then, $User_i$ generates a public and private key pair ($K'^{pu}_{useri}$, $K'^{pr}_{useri}$) corresponding to $PDID_{useri}$.
 After that, $User_i$  generates $M_{A1}$ with $PDID_{useri}$, request for EV charging, and request of proof of the CS's VC. Then, $M_{A1}$ is sent to $CS_j$. $CS_j$ composes and sends a message $M_{A2}$ to $User_i$ with $DID_{CSj}$, its signed VC, and request of proof of the user's VC. 

 \textbf{Step 2:} Upon receiving $M_{A2}$ from the CS, the user verifies the signed VC received from the CS by verifying the signature on it with $K^{pu}_{USP}$ using ECDSA's verify algorithm as mentioned in Algorithm \ref{alg:ECDSAVerify}. Then, credential $cred$, the user's public key $K^{pu}_{useri}$, and nonce $n$ are combined and its hash value is calculated.

 Then, $User_i$ generates a nonce $N_{useri}$ and computes 
$EL_{useri} = {LAI}_{useri}\oplus h({K_{useri}}\parallel{N_{useri}})$ 
 where ${LAI}_{useri}$ is the location identifier of $User_i$. Then, $User_i$ computes  
 ${V_1}= h({PDID}_{useri}\parallel{N_{useri}}\parallel{K_{useri}}\parallel{EL_{useri}})$. Next, the user produces a ZKP of the VC using the \textit{Prove} algorithm with an output $\pi$ as mentioned in Table \ref{table:zkp}. This is the ZKP of the VC the user presents to the CS to prove in Zero-Knowledge that it holds a valid VC. Then, the user composes $M_{A3}:\{ {hashvalue}, {\pi}, {N_{useri}}, {EL_{useri}}, {V_1}\}$ and sends it to $CS_j$. When $CS_j$ receives $M_{A3}$, it verifies $\pi$ using the \textit{Verify} algorithm as mentioned in Table \ref{table:zkp}.

Thus, $User_i$ presents the credential to $CS_j$ using ZKP. The CS is not able to learn anything other than the fact that $User_i$ holds the credential. Once both parties verify that the credentials are legitimate,  $CS_j$ generates a nonce $N_{CSj}$ and sends the request to the USP with its location information $LAI_{CSj}$. For that, $CS_j$  computes  
 ${V_2}= h({DID}_{CSj}\parallel{N_{CSj}}\parallel{K_{CSj}}\parallel{LAI_{CSj}})$. Next, the CS composes a message $M_{A4}:\{M_{A3}, {DID}_{CSj}, {N}_{CSj}, LAI_{CSj}, {V}_{2}\}$ and sends it to the USP.
 
 \textbf{ Step 3:} The location identifier received from $User_i$, $LAI_{useri}$, and from $CS_j$, $LAI_{CSj}$, are compared by the USP. The USP wants to verify that they are the same before proceeding further. This step prevents the location forgery attack. Then, the USP generates a session key $SK$ and a nonce $n^{new}$. After that, $cred$, the user's public key, and \textit{$ n^{new}$} are combined and its hash value, {${hashvalue}^{new}$}, is calculated.
 After that, the USP generates a new VC for $User_i$ based on the new hash value ${hashvalue}^{new}$ that the user can use in the next charging authentication process. Since the VC for each charging event is different, the user cannot be tracked. 
 
 The user's PDID for the next round is stored at the USP as  $PDID^{new}_{useri}= h(PDID_{useri} \parallel K_{useri})$. The session key for $User_i$  is computed as $SK_{useri}= h(PDID_{useri}\parallel N_{useri} \parallel K_{useri}) \oplus SK $. Similarly, the session key for $CS_j$ is computed as $SK_{CSj}= h(DID_{CSj}\parallel N_{CSj} \parallel K_{CSj})\oplus SK$.
 Then, the USP generates a response $V_3$ for $CS_j$ and a response $V_4$ for $User_i$. $V_3$ is computed as  $V_3= h(SK_{CSj} \parallel N_{CSj} \parallel K_{CSj})$ and $V_4$ as $V_4= h(SK_{useri}  \parallel N_{useri} \parallel K_{useri})$. The nonce and the signed VC are XORed with $K_{useri}$ and sent to the user. 
 
 After that, the USP generates a message $M_{A5useri}$ by encrypting $\{({SK}_{useri},{V_4}, {n^{*new}}, {[VC_{useri}]^{*new}_{K^{pr}_{USP}}})\}$ with $K^{pub}_{useri}$.
 Then, the USP generates a message $M_{A5}:\{ M_{A5useri} \parallel({SK}_{CSj}, {V_3})\}$. Then, $M_{A5}$ is sent to the CS.

  \textbf{Step 4:} The CS receives $M_{A5}$ and verifies ${V_3}$. Then, the CS calculates the $SK$ as $SK= h(DID_{CSj}\parallel N_{CSj}\parallel K_{CSj})\oplus SK_{CSj}$ and sends 
  $M_{A6}:M_{A5useri} $ to $User_i$.

  \textbf{Step 5:} After receiving $M_{A6}$, the user decrypts $M_{A5useri}$ with his/her private key $K^{pr}_{useri}$ and verifies the key-hash response ${V_4}$. If the verification is successful, the user decodes $SK$ as $SK= h(PDID_{useri}\parallel N_{useri}\parallel K_{useri}) \oplus SK_{useri}$. The user creates a new ID, $PDID^{new}_{useri}=h(PDID_{useri}\parallel K_{useri})$. The nonce and the signed VC received from the USP are XORed with $K_{useri}$ and stored by the user for next round of authentication. Thus, a session key $SK$ is established among all the parties. The authentication phase is illustrated in Figure \ref{fig:Auth}.

  \subsection{Key Recovery}
  
  \textbf{Step 1:} This step is required for key recovery if the private key is lost. The private key $K^{pr}_{useri}$ is divided into $n$ shares with a $(k,n)$ secret sharing scheme as explained in Section \ref{Prelims}. A random $k-1$ degree polynomial is selected as:

 \begin{equation}
f(x)= K^{pr}_{useri}+a_1(x)+ \cdots +a_{kt-1}(x)^{k-1} \label{step2}.
 \end{equation}

In (\ref{step2}), $a_1, a_2, \cdots, a_{k-1}$ are random polynomial coefficients. After that, the USP evaluates $n$ values as $[K^{pr}_{useri}]_1=f(1)$, $[K^{pr}_{useri}]_2=f(2)$, $\cdots$, $[K^{pr}_{useri}]_n=f(n)$. The shares of the private key calculated as mentioned above can be encrypted and stored in such a way that an adversary cannot correlate the shares.  

 \textbf{Step 2:} With any subset of $k$ of these $n$ values, 
the USP's private key can be calculated as
 \begin{equation}
K^{pr}_{useri} = \Sigma_{j=1}^k f(i_j)  \Pi_{j\neq m} \frac{i_j}{i_j - i_m}.\label{step2_key}
 \end{equation}
 
With the knowledge of $k-1$ of these values, $K^{pr}_{useri}$ cannot be  calculated. Note that during authentication, $M_{A5useri}$ is encrypted with $K^{pu}_{useri}$. The user decrypts it with $K^{pr}_{useri}$. If the private key is lost, it can be reconstituted from the backup using the $(k,n)$ mechanism as mentioned in (\ref{step2_key}).

 \textbf{Remarks:} During authentication, if the synchronization is lost, the user $User_i$ selects one of the unused shadow identities from the set $D_{useri}$, and sends it to the CS in $M_{A1}$. After that, the used shadow identity will be deleted. 
 
 \section{Formal Security Proof}

\subsection{Definitions and assumptions}

The proposed protocol's security is assessed using the Real-Or-Random (RoR) model \cite{abdalla2005password}
Under the RoR model, we show that the proposed protocol can ensure session key security.

First, we will discuss the RoR model briefly. 
The protocol is secure if the established session key cannot be differentiated  from a random string. In this model, security is defined through a series of games played between the participants and an adversary $A$. In our proof, we use imperative properties like collision resistant one-way hash function $h(\cdot)$ which is a pseudo random function and Elliptic Curve Decisional Diffie-Hellman Problem (ECDDHP). In the RoR model, we use the following queries to simulate the attacks.

\textbf{{Queries to model the attacks:}} During the authentication phase, VCs of the user and the CS are verified by each other. The channel through which the authentication messages are exchanged is not secure. An adversary \textit{A} can control the insecure channel between the user and the CS by eavesdropping and modifying the messages sent between them. Let us denote $P^t_{EV}$ as the $t^{th}$ instance of the EV and $P^t_{CS}$ as the $t^{th}$ instance of the CS. The following queries can be used to model these attack scenarios:

\textbf{Execute($P^{t1}_{EV}$, $P^{t2}_{CS}$):} Models attacker \textit{A}’s ability to eavesdrop and intercept the messages communicated between $t_1^{th}$ instance of EV and $t_2^{th}$ instance of CS in a session of the protocol.


\textbf{Reveal($P^{t}$):} Models \textit{A’}s ability to obtain the session key $SK$ established between $P^{t}$ and its partner in a session of the protocol.

\textbf{Test($P^{t}$):} The adversary is allowed to call this query to get the session key, and the output is either the session key $SK$ or a random key based on an unbiased coin or hidden bit $c$. If $c=1$, $P^{t}$ returns $SK$. If $c=0$, $P^{t}$ returns a random number. Otherwise, $P^{t}$ returns null. 

 A one-way hash function $h(\cdot)$ is also modelled as a random oracle. It is accessible to all the parties and $A$.

\textbf{Theorem 1.} Let $A$ be an adversary trying to break the semantic security of the protocol. 
$A$ asks at most $q_h$ hash queries. Let $| {Hash} |$ denote the length of the hash output and let $Adv^{ECDSA}_A$ represent A’s advantage in breaking ECDDHP problem. Then, the advantage of $A$ in breaking the security of the session key in the proposed scheme is $Adv_{A}(t) \leq \frac{(q_h)^2 } { \mid {Hash} \mid}$+  2 $Adv^{ECDSA}_A$ which is negligible.

\textbf{Proof:} Let $G_i$ denote a sequence of games where $i=0,1, \cdots, 2$. Let $Adv_{A,G_i}$ denote $A$'s advantage in the game $G_i$. Let $Success_A^{G_{i}}$ be the event when $A$ correctly guesses the bit $c$ in game $G_i$. Hence, $Adv_{A,G_i} = Pr[ Success_A^{G_{i}}]$. We want the advantage of $A$ to be negligible.

The details of the games $G_i$ where $i=0,1,2$ are given below:

\textbf{Game \textit{$G_0$}:}  This game corresponds to an actual attack by $A$ against the proposed protocol. Since a bit $c$ is selected randomly in $G_0$, we get:

\begin{equation}
Adv_A (t)= \mid {2 Adv_{A,G_0} - 1}\mid. \label{AG0}
\end{equation}

\textbf{Game $G_1$:} In this game, $A$ executes an eavesdropping attack in which $A$ has the capability to intercept all the communicated messages. $A$ calls the $\mathit{Execute}$ query to intercept the transmitted messages $M_{A5}$ and $M_{A6}$. Then, $A$ calls $\mathit{Reveal}$ and $\mathit{Test}$ queries to check if the captured session key is real or random. Session key $SK$ is generated by the USP. $SK_{CSj}$ and $SK_{useri}$, which are transmitted through the messages $M_{A5}$ and $M_{A6}$, respectively, are computed as 
$SK_{useri} = h(PDID_{useri} \parallel N_{useri} \parallel K_{useri}) \oplus SK$ and $SK_{CSj}= h(DID_{CSj}\parallel N_{CSj} \parallel K_{CSj})\oplus SK$. $A$ does not know the secret keys $K_{CSj}$ and $K_{useri}$ to compute $SK$ from $SK_{CSj}$ and $SK_{useri}$. 
To calculate the session key, $A$ should also know the random nonce values $N_{CSj}$ and $N_{useri}$ as well as the pseudo-identity of the user $PDID_{useri}$ and the DID of the charging station $DID_{CSj}$. Hence, eavesdropping the messages $M_{A5}$ and $M_{A6}$ does not increase $A$'s probability to win the game $G_1$. In other words, $G_0$ and $G_1$ are indistinguishable and

\begin{equation}
Adv_{A,G_1}= Adv_{A,G_0}. \label{AG1}
\end{equation}

\textbf{Game $G_2$:} In this game, $A$ makes multiple hash queries and tries to find a message digest collision. A one-way hash function is used for composing the messages $M_{A3}$, $M_{A4}$, $M_{A5}$, and $M_{A6}$. Further, in our protocol, the verifiable credentials change during each iteration of the protocol. They are signed by the private key of the USP using ECDSA. It is a computationally infeasible problem to find the private key from the public key due to intractability property of ECDDHP. Hence, knowing the public key of the USP also does not give any advantage to $A$. $A$ should also know the other required secret parameters in order to compute the session key.
Hence, from the birthday paradox of the hash function and intractability of ECDDHP, we get:

\begin{equation}
\mid Adv_{A,G_1}- Adv_{A,G_2}\mid \leq \frac{(q_h)^2 } {2 \mid {Hash} \mid}+Adv^{ECDSA}_A(t).  \label{AG2}
\end{equation}

After all the above games are executed, $A$ guesses the bit $c$ and calls $Test$ query to win the game. Then, we get the following:

\begin{equation}
Adv_{A,G_2} = \frac {1}{2}.  \label{AG2_last}
\end{equation}

Combining (\ref{AG0}) and (\ref{AG1}), we get the following:
\begin{eqnarray}  
\label{final1}
\frac{1}{2} Adv_{A}(t) &=& \mid { Adv_{A,G_0} - \frac{1}{2}}\mid \nonumber \\
                              &=& \mid { Adv_{A,G_1} - \frac{1}{2}}\mid.
\end{eqnarray}

From (\ref{AG2}), (\ref{AG2_last}), and (\ref{final1}), we have: 

\begin{eqnarray} 
\label{final2}
\frac{1}{2} Adv_{A}(t) &=& \mid { Adv_{A,G_1} - \frac{1}{2}}\mid \nonumber \\
    &=& \mid Adv_{A,G_1} - Adv_{A,G_2}\mid \nonumber \\
    &\leq& \frac{(q_h)^2 } {2 \mid {Hash} \mid}+Adv^{ECDSA}_A(t).
\end{eqnarray}

By multiplying both sides of (\ref{final2}), by 2, we get:

\begin{equation}
Adv_{A}(t) \leq \frac{(q_h)^2 } { \mid {Hash} \mid}+2 Adv^{ECDSA}_A(t).
\end{equation}

Hence, this shows that the proposed protocol ensures session key security.
\hfill $\blacksquare$

\section {Informal Security Analysis}

 In this section, we demonstrate how the proposed protocol achieves some of the important security properties for EV charging. We also discuss how the proposed protocol prevents the threats mentioned in the adversary model.

\subsection {Security Properties} \label{SPro}
The key security features of the scheme are discussed below.

\textbf{User-empowerment}: The proposed scheme is based on DID. The users create and manage their IDs. Their IDs are not issued by any centralised issuing authority. Hence, the users have control over their IDs which empowers them.

\textbf{Biometric Authentication}: The user inputs his/her biometrics (e.g., fingerprint) $\beta_i$ and password $psw_i$ to access the mobile device $MD_i$. Then, $MD_i$ computes  $\delta'_{i} = h(\beta_i \parallel psw_i)$ and compares $\delta'_i$ with $\delta_i$ to verify $User_i$'s legitimacy. This biometric verification at the mobile device is to ensure the user's legitimacy. The remianing steps of authentication will be carried out only if the biometric verification is successful. Even if an adversary gets the mobile device, he/she cannot use it for EV charging authentication as the biometric verification will fail.

\textbf{Mutual Authentication}: The user presents his/her VC to $CS_j$. The charging station presents its VCs to $User_i$. The user verifies the VC of $CS_j$. Then, the user presents ZKP of his/her VC to the CS and the CS verifies it. Only a legitimate user and legitimate CS can provide a valid VC, signed with the private key of the USP $K^{pr}_{USP}$. Thus, both the parties verify VC of the other party and the proposed protocol achieves mutual authentication.

  \textbf{Anonymity}: Users reveal their real identities only to the USP. To charge the vehicle, the user uses a pseudo-ID $PDID_{useri}$ and ZKP of a VC signed by the USP that the charging station verifies. The user's real identity is not revealed during the charging process. Thus, the proposed scheme maintains anonymity of the users.
  
    \begin{table*}[t]
 		\begin{center}
		\caption{COMPARISON BASED ON SECURITY FEATURES }
		
		\scalebox{1.0}{

		\begin{tabular}{|c|c|c|c|c|c|c|c|c|}

				\hline
				\textbf{Scheme} & \textbf{SF1} & \textbf{SF2} &	\textbf{SF3}& \textbf{SF4} & \textbf{SF5}&
				\textbf{SF6} & 	\textbf{SF7} & 	\textbf{SF8}\\

				\hline
				
										Roman et al.\cite {roman2019pairing} & Yes & Yes  &Yes &  Yes & Yes &  Yes & No & No\\
			
			\hline
			
			Gope et al.\cite{gope2019efficient} & Yes	 &  Yes & Yes & Yes  & Yes & Yes & No & No\\
			
			\hline	
				
			Zhang et al. \cite{zhang2021efficient}   & Yes & Yes  &Yes &  Yes & Yes &  No & No & No\\
			
			\hline
			
			Bansal et al.\cite {Bansal9018072} & Yes & Yes  &Yes &  Yes & Yes &  No & No & No\\
			
			\hline
			
						 Saxena et al.\cite {saxena2015lightweight} & Yes & Yes  &Yes &  Yes & Yes &  No & No & No\\
			
			\hline

			$Proposed$ $Scheme$   & Yes & Yes  &Yes &  Yes & Yes &  Yes & Yes & Yes\\

			\hline
		\multicolumn{9}{|c|}{\textbf{SF1:} Privacy of the user; \textbf{SF2:} Anonymity of the user; \textbf{SF3:} Session key; 
			 } \\
			\hline
			\multicolumn{9}{|c|}{\textbf{SF4:} Protection against replay attacks; \textbf{SF5:} Mutual authentication;  
			 } \\

            \hline
			\multicolumn{9}{|c|}{ \textbf{SF6:} Protection against location forgery attacks; 
			 } \\
			  \hline
			\multicolumn{9}{|c|}{ \textbf{SF7:} Non-repudiation; \textbf{SF8:} User-empowerment
			 } \\

           \hline
 			
			\end{tabular}
			}
			
			\label{table:comparison}
		\end{center}

	\end{table*}

\begin{table*}[t]

		\begin{center}
		\caption{COMPARISON BASED ON COMPUTATION COST}
		
		\scalebox{1.}{

		\begin{tabular}{|p {2.5cm}|p {5.9 cm}|p {5.9 cm}|}

				\hline
				\textbf{Scheme} & \textbf{User's Device} & \textbf{USP/CS}  \\
							\hline
			
							Roman et al.\cite {roman2019pairing}  & $3M+B$ = 26.13  ms & $16M+3B$ = 86.84 ms  \\
			
			\hline

			Gope et al.\cite{gope2019efficient} & 6$H$ = 6.84 ms	 &  8$H$ = 9.12 ms \\
			\hline	
				
				Zhang et al. \cite{zhang2021efficient}& 
			$3M+EXP+5H$ = 24.58 ms & $3M+B+4H$ = 30.69 ms  \\
			
			\hline

			Bansal et al.\cite {Bansal9018072}& $1P+2MAC+4EXP+2S+2M$ = 31.874 ms & $1P+6MAC+4S+6EXP+4M$ =  65.814 ms  \\
			
			\hline

		Saxena et al.\cite {saxena2015lightweight}& $1M+3EXP+1B$ =  25.13 ms & $5M+8EXP+1B$ =  61.89 ms  \\
			
			\hline

			$Proposed$ $Scheme$   & $6H+ Verify_{ECDSA}$ = 28.06 ms & $8H+ Sign_{ECDSA}+ Verify_{ECDSA}$ = 45.95 ms   \\
				\hline
		\multicolumn{3}{|c|}{ \textbf{$M$:}  Multiplication Operation =5.24 ms; \textbf{$MP$:}  Multiplication Point Operation = 3.21 ms; \textbf{$P$:}  PUF Operation; \textbf{$H$:} Hash Operation =1.14 ms;
			 } \\
			\hline
			\multicolumn{3}{|c|}{ \textbf{$EXP$:} Modular Exponential Operation =3.16 ms; \textbf{$S$:} Symmetric Encryption/Decryption Operation =0.17 ms; \textbf{$MAC$:}  MAC Operation =1.3 ms;
			 } \\

			 \hline
			\multicolumn{3}{|c|}{ \textbf{$B$:} Bilinear Pairing Operation =10.41 ms; \textbf{$Sign_{ECDSA}$:} ECDSA Signing = 15.61 ms;  \textbf{$Verify_{ECDSA}$:} ECDSA Signature Verification =21.22 ms } \\

           \hline

			\end{tabular}
			}
			
			\label{table:perfcomparison}
		\end{center}

	\end{table*}

\textbf{Unlinkability}: During authentication, the user provides pseudo ID $PDID_{useri}$ and ZKP of a verifiable credential signed by the USP that the CS verifies. For two consecutive sessions $x$ and $x+1$, $PDID^x_{useri} \neq PDID^{x+1}_{useri}$. Thus, the identities of the users are unlinkable.

\textbf{Privacy of the User}: To charge the EVs, the users request service with their pseudo-identities. The user also presents a ZKP of verifiable credential signed by the USP ($[VC_{useri}]_{K^{pr}_{USP}}$) that the CS verifies. By using ZKP, the CS is not able to learn anything about the user. The ZKP provides proof about the VC to show that the user is a valid registered user but without disclosing anything about the user. Further, the pseudo-identities are different in two consecutive sesssions. Only the trusted party (the USP)  knows the actual identity of the user. The pseudo identity of the user and the VC are changed during each session. For two consecutive sessions $x$ and $x+1$, $PDID^x_{useri} \neq PDID^{x+1}_{useri}$. An adversary will not be able to link a user's real identity with his/her pseudo ID or VC, and will not be able to track the user's trajectory or routines. Thus, the proposed protocol provides privacy.

\textbf{Accountability}: During registration, the USP verifies the VC of the digital identity that is signed by a trusted organization such as a government agency e.g., ${[VC_{DigIDuseri}]_{K^{pr}_{Gov}}}$. Thus, the USP confirms the legitimacy of the user. This legitimacy check of the user by the USP ensures accountability.

\textbf{Session Key Agreement}: During authentication, a session key $SK$ is generated by the USP. 
The USP sends $SK_{CSj}= h(DID_{CSj}\parallel N_{CSj} \parallel K_{CSj})\oplus SK$
and $V_3= h(SK_{CSj} \parallel N_{CSj} \parallel K_{CSj})$ to the CS through $M_{A5}$. When the CS receives $M_{A5}$, $SK$ is calculated as $SK= h(DID_{CSj}\parallel N_{CSj} \parallel K_{CSj})\oplus SK_{CSj}$. Similarly, the USP sends $SK_{useri}= h(PDID_{useri}\parallel N_{useri} \parallel K_{useri})\oplus SK$
and $V_4= h(SK_{useri} \parallel N_{useri} \parallel K_{useri})$ to the user. When the user receives $M_{A6}$, $SK$ is calculated as $SK= h(PDID_{useri}\parallel N_{useri} \parallel K_{useri})\oplus SK_{useri}$. Thus, there is a session key agreement among the participants of the proposed protocol.

\textbf{Non-Repudiation}: After signing a statement with its private key, a party cannot deny having signed it (i.e., non-repudiation). The VC signing and verification is based on asymmetric cryptography. The USP signs the VC for the user and the CS with its private key ($[VC_{useri}]_{K^{pr}_{USP}}$). The private key $K^{pr}_{USP}$ used to sign the VC is only known to the USP. This ensures non-repudiation.

\textbf{Protection Against Impersonation Attacks}: To impersonate a legitimate user $User_i$, the adversary needs to send $M_{A3}:\{{hashvalue}, \pi, {N}_{useri},{EL_{useri}},{V}_{1}\}$ to the CS. Only the user knows $[VC_{useri}]_{K^{pr}_{USP}}$ and the parameter $K_{useri}$ to compute $EL_{useri} = {LAI}_{useri}\oplus h({K_{useri}}\parallel{N_{useri}})$ and ${V_1}= h({PDID}_{useri}\parallel{N_{useri}}\parallel{K_{useri}}\parallel{EL_{useri}})$ which are used to compose $M_{A3}$. Hence, an adversary's attempt to impersonate a registered user will not be successful. Similarly, to impersonate a registered CS, the adversary needs to generate $M_{A4}:\{M_{A3}, {DID}_{CSj},{N}_{CSj}, LAI_{CSj}, {V}_{2}\}$. Only the CS knows the secret parameter $K_{csj}$ to compute ${V_2}= h({DID}_{CSj}\parallel{N_{CSj}}\parallel{K_{csj}}\parallel{LAI_{CSj}})$. To impersonate the USP, the attacker needs to have the knowledge of $K_{csj}$ and $K_{useri}$ to generate $V_3= h(SK_{CSj} \parallel N_{CSj} \parallel K_{CSj})$ and $V_4$ as $V_4= h(SK_{useri} \parallel N_{useri} \parallel K_{useri})$. The adversary does not have the knowledge of $K_{csj}$ and $K_{useri}$ to generate valid key-hash responses. Thus, the proposed protocol is robust against impersonation attacks.

\textbf{Protection Against Location Forgery Attacks}: The EV charging price depends on the location of the CS. A dishonest user or a dishonest CS may provide a false location identifier to the USP. When the USP receives message $M_{A4}$, it decodes ${LAI}_{useri}$ from $EL_{useri}$ and compares ${LAI}_{useri}$ with $LAI_{CSj}$. If the comparison of location identities fails, the authentication process will be terminated. Thus, the proposed authentication protocol prevents location forgery attacks.

\textbf{Protection Against Replay Attacks}: An adversary may capture the messages transmitted between different entities and may replay it later to get authenticated. However, in the proposed scheme, $[VC_{useri}]_{K^{pr}_{USP}}$ and ${N}_{useri}$ in $M_{A3}:\{{hashvalue}, \pi, {N}_{useri}, {EL_{useri}}, {V}_{1}\}$ are not repeated in two sessions. Similarly, the parameter $N_{CSj}$ in $M_{A4}:\{M_{A3}, {DID}_{CSj}, {N}_{CSj}, LAI_{CSj}, {V}_{2}\}$ is not repeated. Hence, an adversary's attempt to capture and replay the messages $M_{A3}$ and $M_{A4}$ to get authenticated will not be successful. Similarly, the key-hash responses $V_3$ and $V_4$ from the USP depend on the values $N_{CSj}$ and $N_{useri}$ are not repeated. This ensures that the adversary's attempt to replay the captured messages $M_{A5}$ and $M_{A6}$ from the USP will not be successful as well since they are composed of $V_3$ and $V_4$, respectively. Thus, the proposed scheme prevents replay attacks.

 \section{Performance Analysis and Comparison} 

 This section compares the proposed scheme's security features with that of other similar solutions in the literature. Then, we evaluate the proposed protocol's computation cost. After that, we  perform a comparison of the results with that of other similar protocols. A comparison of the security features is given in Table \ref{table:comparison}. The proposed scheme's main distinguishing feature from other works is user-empowerment. For EV charging in the smart grid, none of the existing authentication schemes leveraged DID to empower the users. Also, while most similar works ensure privacy and anonymity for the user, they do not provide non-repudiation. The proposed scheme provides non-repudiation. The proposed scheme prevents location forgery attacks. Most of the other schemes do not protect against location forgery attacks. Hence, the proposed scheme achieves all the essential security requirements together with user empowerment.
 
 \begin{figure}
\centering
\begin{minipage}{.5\textwidth}
\centering
\includegraphics[width=\linewidth]{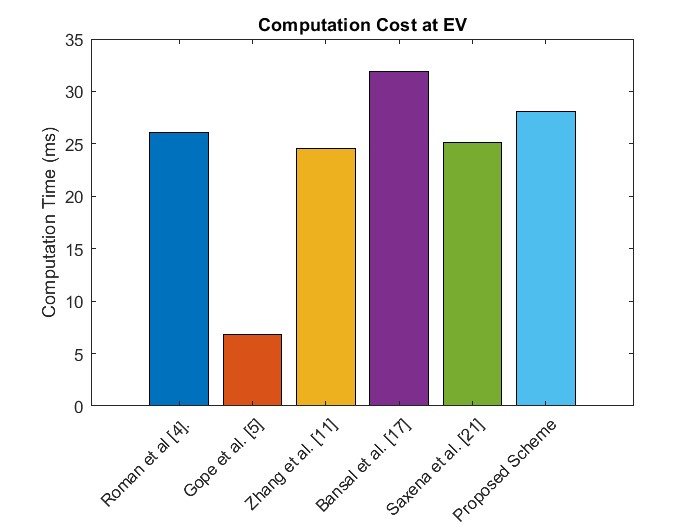}
\caption{Comparison of computation cost at EV.}
\label{fig:userdevice}
\end{minipage}\hfill
\begin{minipage}{.5\textwidth}
\centering
\includegraphics[width=\linewidth]{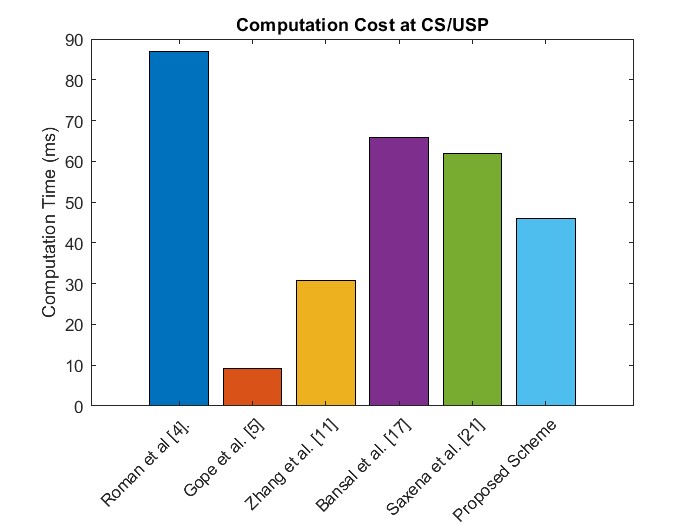}
\caption{Comparison of computation cost at CS/USP.}
\label{fig:csusp}
\end{minipage}\hfill

\end{figure}

\begin{figure}

\centering
\begin{minipage}{.5\textwidth}
\centering
\includegraphics[width=\linewidth]{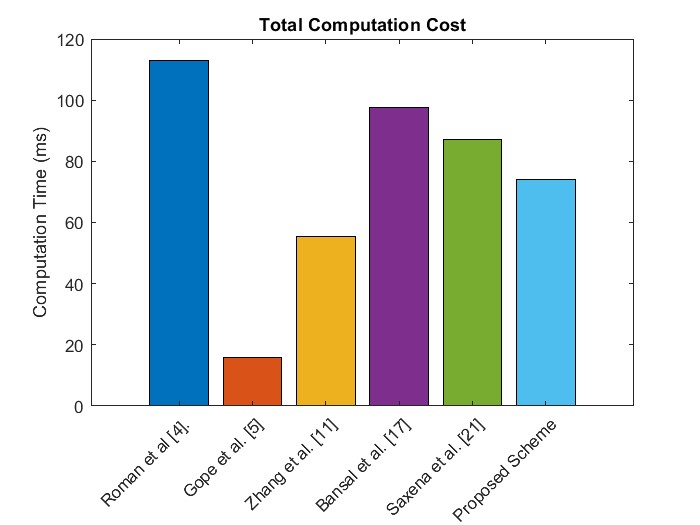}
\caption{Comparison of total computation cost.}
\label{fig:total}
\end{minipage} \hfill

\begin{minipage}{.5\textwidth}
\centering
\includegraphics[width=\linewidth]{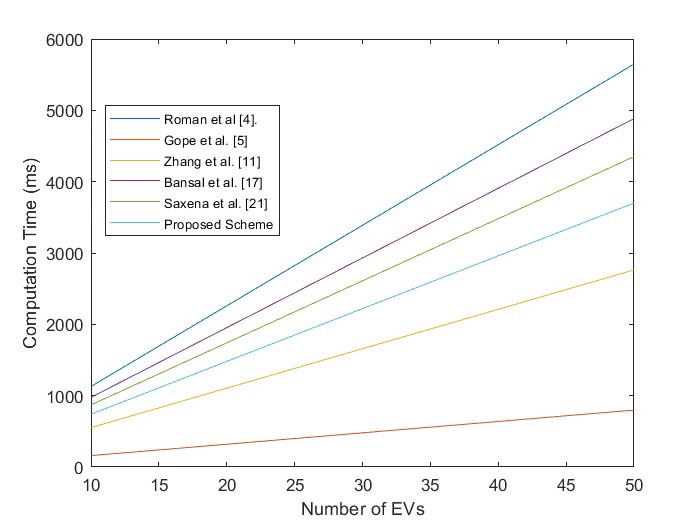}
\caption{Computation cost as a function of the number of EVs.}
\label{fig:scale}
\end{minipage}\hfill

\end{figure}

   Now, we evaluate the computation cost of the proposed scheme. The registration process is executed only once. Hence, the computation cost during the registration phase is not significant for the performance of the proposed scheme. However, the authentication process is executed each time the user wants to charge the vehicle. Hence, we discuss the computation cost during the authentication phase of the protocol in this section while omitting the computation cost during the initial registration.
 Since the time taken by the XOR operation and concatenation operation is negligible, the time taken by these operations is not considered when evaluating the computation cost of the protocols. 
 To simulate the experiments, we employ a personal computer with an Intel Core i5 CPU, 2.90 GHz clock, and 4 GB of RAM as the CS/USP. The time taken by various cryptographic operations is given in Table \ref{table:perfcomparison}. Then, we compare the computational cost of the proposed scheme 
 with other schemes. The performance comparison based on computation cost is presented in Table \ref{table:perfcomparison}.

To compare the computation time taken by the user's device in different schemes during authentication, the computation time is calculated and is plotted in Figure \ref{fig:userdevice}. Similarly, the computation time taken by the CS/USP and the total time taken during authentication in different schemes is plotted in Figure \ref{fig:csusp} and Figure \ref{fig:total}, respectively. We now look at how the computation time increases when the number of EVs is increased. The computation cost as a function of the number of EVs is plotted in Figure \ref{fig:scale}. When a user's VC is verified, the USP sends the user a new VC for the next round. This step is required to provide unlinkability between the VCs in consecutive sessions. The computation cost at the USP in the proposed scheme is slightly higher than that in some other schemes, since generating a new VC adds to the computation cost at USP in the proposed scheme. Considering the fact that the USP has sufficient resources to do the computation and the security properties the proposed scheme offers, the computation cost at the USP is reasonable. As a result, we can conclude that the proposed scheme has a reasonable computation time.

 \section{Conclusion}
 
In this paper, we presented a DID, VC, and ZKP-based authenticated key exchange protocol for EV charging in smart grid that allows users to create and control their IDs. They can charge their electtric vehicles without revealing their identities. The user presents the ZKP of the VC to prove his/her legitimacy before charging the vehicle. By making use of ZPK, the CS is not able to learn anything about the user in this protocol. The proposed protocol also provides an option to recover the private key of the user in the event it is lost. The key recovery mechanism makes the proposed protocol resilient to accidental loss of a private key. We provided informal and formal security analyses to show that the proposed mechanism is robust. The proposed mechanism achieves security properties such as session 
key security, mutual authentication, privacy, anonymity, unlinkability, and protection against many attacks. We compared the proposed scheme with other similar works regarding performance and security properties. Our analysis reveals that the proposed scheme provides all major security features and privacy for the user at a reasonable computational cost.

\bibliographystyle{IEEEtran}
\bibliography{references}

\end{document}